# Fundamental properties alkali-intercalated bilayer graphene nanoribbons

Thi My Duyen Huynh[1], Guo-Song Hung[2], Godfrey Gumbs[3], Ngoc Thanh Thuy Tran[4*]

[1]Department of Physics, National Cheng Kung University, Tainan 701, Taiwan

[2]Department of Materials Science and Engineering, National Cheng Kung University, Tainan 701, Taiwan

[3]Department of Physics and Astronomy, Hunter College of the City University of New York, New York 10065, United States.

[4]Hierachical Green-Energy Materials (Hi-GEM) Research Center, National Cheng Kung University, Tainan 701, Taiwan

*Email of the corresponding author: tranntt@phys.ncku.edu.tw

**Abstract**

Along with the inherent remarkable properties of graphene, adatom-intercalated graphene-related systems are expected to exhibit tunable electronic properties. The metal-based atoms could provide multi-orbital hybridizations with the out-of-plane π-bondings on the carbon honeycomb lattice, which dominates the fundamental properties of chemisorption systems. In this work, using the first-principles calculations, the feature-rich properties of alkali-metal intercalated graphene nanoribbons (GNRs) are investigated, including edge passivation, stacking configurations, intercalation sites, stability, charge density distribution, magnetic configuration, and electronic properties. There exists a transformation from finite gap semiconducting to metallic behaviors, indicating enhanced electrical conductivity. They arise from the cooperative or competitive relations among the significant chemical bonds, finite-size quantum confinement, edge structure, and stacking order. Moreover, the decoration of edge structures with hydrogen and oxygen atoms is considered to provide more information about the stability and magnetization due to the ribbons' effect. These findings will be helpful for experimental fabrications and measurements for further investigation of GNRs-based materials.

## 1. Introduction

Graphene nanoribbons (GNRs), a one-dimensional (1D) narrow strip of graphene, have inspired a lot of studies due to their easy synthesis, ability to open band gaps, and other remarkable properties[1-4]. Through the 1D quantum confinement effects of GNR, the essential properties can be enhanced significantly, which could help overcome the limitations of 2D graphene with its zero-gap electronic structure[5]. Nanoribbon width and edge structure are crucial to the significant characteristics of GNRs, especially in bandgap engineering and magnetic configuration. Edges of GNRs can be characterized in two common types, namely armchair and zigzag (AGNRs and ZGNRs). The former belongs to non-magnetic semiconductors, while the latter are antiferromagnetic (AFM) semiconductors[6]. GNRs have been successfully synthesized under the top-down and bottom-up schemes including oxidation reaction[7, 8], chemical vapor deposition[9-11], and unzipping carbon nanotubes[12, 13]. However, roughness occurring at the ribbon edges during the synthesis process can reduce the mobility of GNRs because of the edge-scattering effects[14, 15]. To resolve this problem, adatoms passivation at the edge is reported as an effective method[5, 15, 16]. The chemical bonds between adatoms and carbons can reconstruct edge structures and thereby drastically alter the electronic features. This study focuses on two typical adatom passivation types, H[5, 15, 16], and O[17-19], which are promising candidates in experimental and theoretical studies, respectively.

GNRs' properties could be modulated by layered structures[20], chemical or physical modifications[21-25] such as doping[21, 26]/ adsorbing[27, 28], functionalizing[24, 25], or applying external fields[29, 30] for expanding the range of applications. It is worth mentioning that few-layer GNRs sharply contrast with monolayer systems in essential features, mainly owing to the various stacking configurations and interlayer/intralayer spin distributions[31, 32]. The calculated results show that stacking configurations, especially AA and AB ones, reveal significant effects on electronic properties[32-34]. However, there are lacking studies regarding electronic properties, magnetic configurations, and geometric structures of O-passivated layered GNRs as compared to the H-GNRs one. Moreover, despite H-GNRs are well studied theoretically, predictions are not consistent with one another related to the existence of magnetism and the planar/non-planar geometry[6, 35, 36]. Thus, more complete and accurate results are required to clearly identify the stacking-induced diverse behaviors in O- and H- passivated bilayer GNRs.

Along with structural alternation, chemical modification on GNRs is the most effective strategy in creating dramatic changes between the semiconducting and metallic behaviors. For layered systems, adatom intercalated GNRs have attracted more interest and concerns[37-40] as promising for interconnect[41, 42] materials. Among intercalated atoms, alkali metals have received long-standing attention for intercalation[43-46] because of their simple electronic configuration and potential applications[42] in energy storage. The alkali-intercalated possibility has been demonstrated in bilayer graphene[10, 47] and $MoS_2$, indicating high charge transfer from the alkali atoms to the host material[48]. Besides, varying the concentration of alkali atoms intercalation could adjust the work function of graphene on Cu(111)[49] and the mean height of the carbon layer of graphene on Ir(111)[44] explained through charge transfer. Moreover, alkali intercalated can be used to manipulate the electronic properties of graphene[45, 46] that could realize superconducting behavior[50] in bilayer graphene. Hence, alkali intercalated graphene systems suggest the possible and potential ways for model study and applications that are expected to tune essential properties of bilayer GNRs. The single-orbitals of alkali atoms (Li/Na) are expected to enhance the conductivity that might have capable applications as interconnect materials or anode materials for Li/Na-ion batteries.

In this study, the fundamental properties of alkali-intercalated bilayer GNRs are systematically investigated by means of the first-principles calculations. Both oxygen and hydrogen passivation are taken into account for comparison. The calculations focus on the formation energies, the adatom-carbon bond lengths, the optimal intercalated position, the adatom-dominated bands, the orbital projected density of states (PDOS), the charge density distributions, and the significant competitions of the carbon ribbon's edge and the adatoms passivation/intercalation in spin configurations. The essential properties arising from distinct types stacking layers, ribbon edge structures, as well as passivated and intercalated adatoms will be discussed in detail.

## 2. Computational Method

The first-principles calculations were performed with the use of the Vienna ab initio simulation software package (VASP)[51]. The projector augmented wave method was implemented to evaluate the electron-ion interactions, in which the electron-electron Coulomb interactions belong to the many-particle exchange and correlation energies under the Perdew-Burke-Ernzerhof (PBE)[52] generalized gradient approximation. The spin configurations are considered to explore the chemical absorption effects on the magnetic properties. A vacuum of 15 Å was set to suppress the van der Waals (vdW) interactions between neighboring slabs. Besides, a DFT-D3 vdW interaction was applied for examining interaction between two layers[53]. A plane-wave basis set with a maximum energy cut-off of 500 eV is available in the calculations of Bloch wave functions. For the criterion of energy and force, all atomic coordinates were relaxed until the force on each atom is less than 0.001 eV/ Å and the energy convergence was set at $10^{-5}$ eV. The k-points of 15x1x1 in Gamma symmetry and high-symmetric Γ-centered Brillouin zone grid were sampled for structural optimization and band energy calculations, respectively.

## 3. Results and discussion

### 3.1 Monolayer and bilayer GNRs

Edge passivation could control electronic properties of GNRs, which is used to precisely fabricate the ribbon structure. Among passivated atoms, hydrogen is reportedly the best theoretically; however, oxygen is highly considered in experiments[5, 15, 16, 18, 19, 35, 36, 54-57]. In order to compare their stability, the formation energies calculation is used, in which a smaller value indicates more favorable structures than others. The formation energy $E_f$ of GNRs with hydrogen and oxygen edge passivation can be obtained as[58-60]

$$E_f = \left[E_{sys} - \left(n_C * E_C + \frac{n_{O/H}}{2} * E_{O_2/H_2}\right)\right]/2L \qquad (1)$$

where $E_{sys}$ is the total energy of GNR with oxygen or hydrogen termination; $E_C$ and $E_{O_2/H_2}$ are energy per atom of bulk AB graphite and oxygen or hydrogen molecules, respectively; $n_C$ and $n_{O/H}$ are the numbers of carbon and oxygen or hydrogen atoms, respectively; and L is the length of the edge. According to $E_f$ results for monolayer cases (see Fig.S1 for structural optimization) shown in Table 1, oxygen passivation exhibits more stable structures than that of H-cases based on smaller formation energies that is in good agreement with experimental studies[61-63]. However, hydrogen-passivated edges are theoretically reported to prevent the edge rearrangement due to passivating dangling bonds[5, 15, 16]. A hydrogen atom can cause further transformation of a $sp^2$ hybridized carbon atom at the edge to a $sp^3$ hybridized atom, resulting in hydrogen-saturated edges. Therefore, both kinds of passivation are considered for further constructing bilayer systems.

Following the arrangement of stacking in graphite[64-67], O- and H- passivated bilayer GNRs can be stacked in three types, namely AA, $AB_\alpha$ and $AB_\beta$, as shown in Fig.1 and Fig.S2, respectively. As described in these figures, AA is formed by arranging the section layer is directly on the top of the first layer while $AB_\alpha$ and $AB_\beta$ are created by translating the second layer one C-C bond length along the armchair and zigzag direction, respectively. Table 1 illustrates the formation energies of oxygen and hydrogen passivation bilayer GNRs, where oxygen passivation dominates in terms of structural stability. The more stable phases of ZGNR and AGNR are, respectively, AA and $AB_\beta$ for H-passivated cases and $AB_\alpha$ and $AB_\beta$ for O-passivated cases. Therefore, these structures are further utilized for alkali intercalated systems. Additionally, the bandgap and interlayer distance are listed in this table, indicating the effect of stacking. For example, AA and $AB_\beta$ exhibit metallic behavior while $AB_\alpha$ shows a very small gap structure in the H-ZGNR system. Also, AGNRs display larger-bandgap compared to ZGNRs both H- and O-passivated cases (see Table 1). Bilayer GNRs reveal various stackings leading to enlarge geometry and electronic properties of GNRs.

**Table 1.** Formation energies, bandgap, interlayer distance of H-/O-passivated monolayer and bilayer GNRs with AA, $AB_\alpha$ and $AB_\beta$ stacked structures.

|  |  | Monolayer |  | Bilayer |  |  |  |  |  |
|---|---|---|---|---|---|---|---|---|---|
|  |  | H-passivated | O-passivated | H-passivated | | | O-passivated | | |
|  |  |  |  | AA | $AB_\alpha$ | $AB_\beta$ | AA | $AB_\alpha$ | $AB_\beta$ |
| ZGNRs | | | | | | | | | |
| $E_f$ (eV) | | -0.8075 | -2.4722 | -1.7149 | -1.7048 | -1.6916 | -4.9482 | -4.9815 | -4.9768 |
| $E_g$ (eV) | | 0.4 | 0.175 | M | 0.035 | M | 0.013 | M | 0.11 |
| d (Å) | $d_1$ | -- | -- | 3.017 | 3.383 | 3.688 | 4.081 | 3.38 | 3.56 |
|  | $d_2$ | -- | -- | 3.554 | 3.352 | 3.51 | 3.75 | 3.438 | 3.452 |
| AGNRs | | | | | | | | | |
| $E_f$ (eV) | | -1.528 | -2.8373 | -2.6582 | -2.7005 | -2.7009 | -5.6962 | -5.7143 | -5.7437 |
| $E_g$ (eV) | | 0.8 | M | 0.531 | 0.49 | 0.6 | M | 0.05 | M |
| d (Å) | $d_1$ | -- | -- | 3.977 | 3.637 | 3.765 | 3.97 | 3.57 | 3.326 |
|  | $d_2$ | -- | -- | 3.95 | 3.692 | 3.404 | 3.683 | 3.462 | 3.52 |

M: metallic

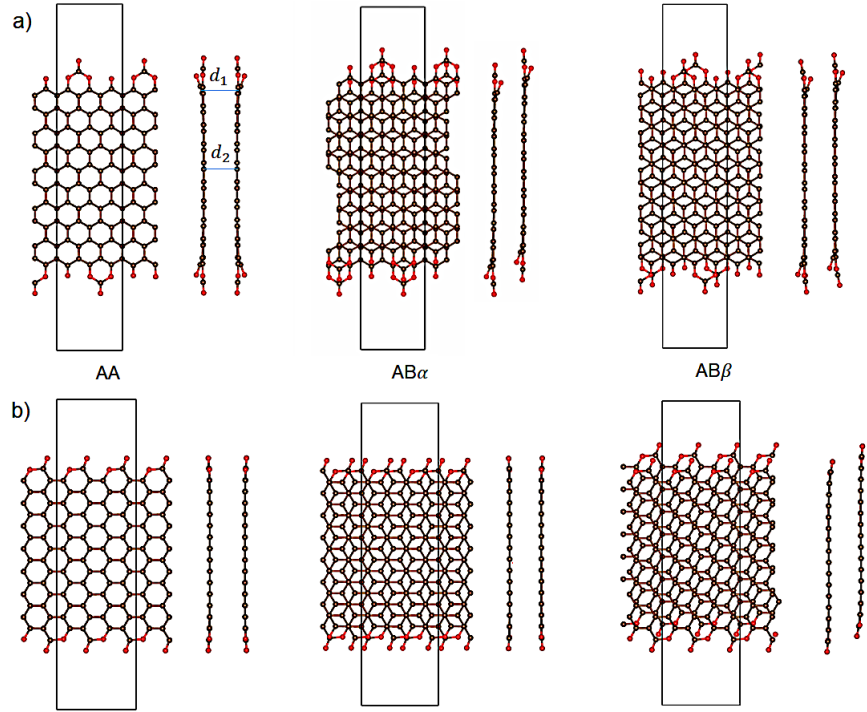

**Fig. 1** The top and side views of bilayer O-passivated (a) ZGNRs and (b) AGNRs with considering three kinds of stacking AA, AB$_\alpha$ and AB$_\beta$, respectively. $d_1$ and $d_2$ describe interlayer distances of two layers near the edge and in the middle of the ribbon, respectively.

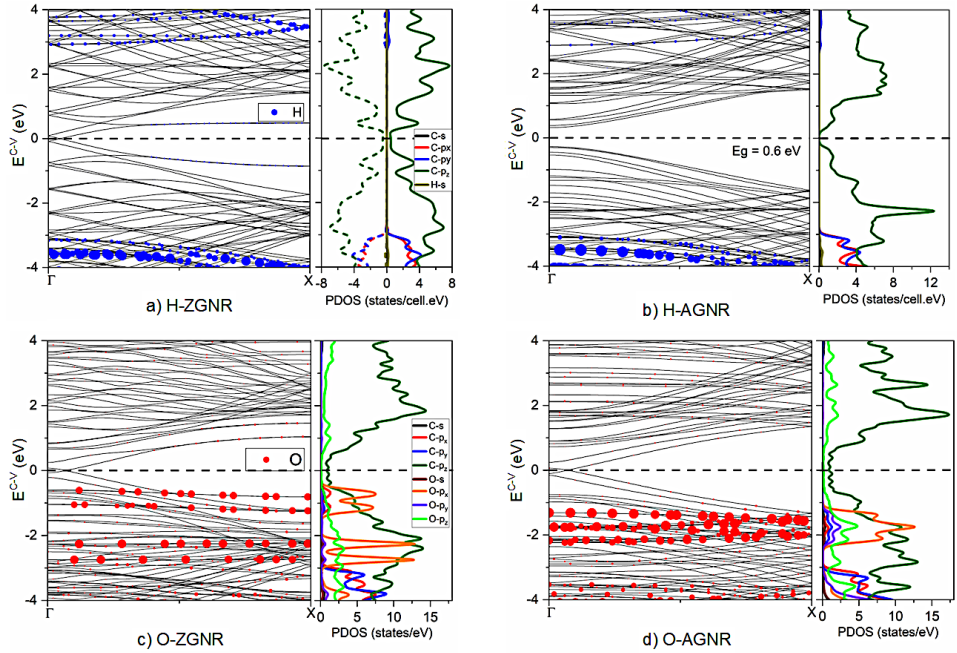

**Fig. 2** Band energy structures and corresponding orbital-projected density of states (DOS) of H-passivated (a) ZGNR_AA, (b) AGNR_AB$_\beta$ and O-passivated (c) ZGNR_AB$_\alpha$, (d) AGNR_AB$_\beta$. Blue

and red circles represent hydrogen and oxygen atom dominance, respectively. Solid and dash lines respectively describe spin up and down orbital density of states.

Electronic properties of GNRs are mainly determined by the edge structure, stackings, and quantum confinement effect. It is worth mentioning that bilayer GNRs are quite different from monolayer one (see Fig.S3 in **Support Information**) in terms of energy dispersion, band overlap/bandgap, and van Hove singularities (vHs) in DOS, as a result of stacking effect. Monolayer GNRs exhibit the special 1D band structures owning to the honeycomb lattice symmetry, finite-size quantum confinement, and edge structure. ZGNR (Fig.S3a) has a pair of partially flat valence and conduction bands nearest to the Fermi level ($E_F$) forming a direct gap of 0.4 eV, corresponding to wave functions localized at the zigzag boundaries due to the strong competition between quantum confinement and zigzag edge spin configuration (discussed later). For AGNRs, most of the energy dispersions are parabolic bands, while a few of them are partially flat ones. The low-lying electronic states within $\pm 2$ eV and the deeper ones are, respectively, contributed by the $\pi$ bonds of parallel $2p_z$ orbitals and the $\sigma$ bonds of $(2s, 2p_x, 2p_y)$ orbitals as described in the orbital-projected DOS. The electronic structures of H- and O-passivated bilayer GNRs (Fig.2) are systematically examined to realize the intercalation effect. The low-lying bands are composed of parabolic/linear and partially flat dispersions. Furthermore, atomic edge illustrates distinct features in band structures, demonstrated in band gap and band overlap, in which H-AGNR (Fig.2b) belongs to direct band gap of 0.6 eV, whereas a pair of linear bands intersecting at $E_F$ is found in H-ZGNR (Fig.2a) and O-GNR (Fig.2c, d). Obviously, oxygen atoms perform a strong passivation effect, indicated in their significant dominance near the Fermi level while hydrogen atoms only exhibit weakly contribute at the middle-energy $|E^{c,v}| > 3$. There exist the $sp^2s$ hybridizations in the C-H bonds, since there are peaks in DOS dominated simultaneously by H-1s and C-$(2s, 2p_x, 2p_y)$ orbital. On the other hand, the strong C-O bonds reveals the $2p_z$-$2p_z$ and $(2s, 2p_x, 2p_y)$- $(2s, 2p_x, 2p_y)$ orbital hybridizations. The translation from $\pi$ to co-exist $\pi$- $\sigma$ bonds in the valence bands near the Fermi level between H- and O-case corresponding to the strong distribution of C-$p_z$ and C-$p_z$, C-$p_x$, C-$p_y$, O-$p_z$ orbitals in PDOS, respectively, further reveals the effect of atomic edge structure.

**3.2 Alkali adatom-intercalated bilayer GNRs**

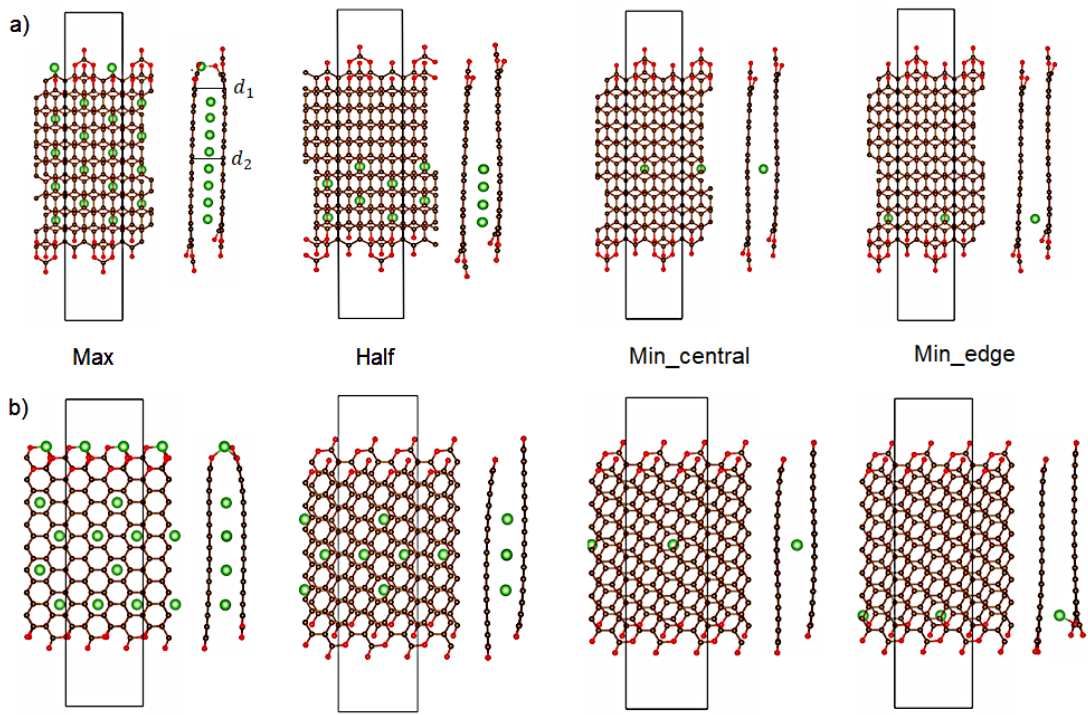

**Fig. 3** Structural optimization of Alkali-intercalated GNRs in O-passivated (a) zigzag and (b) armchair edges with max, half, min_central, and min_edge intercalations. Green circles represent alkali atoms intercalated bilayer GNRs.

Alkali adatom intercalate bilayer GNRs can create an unusual geometric structure, being sensitive to the Alkali (Li, Na) distribution and concentration. As discussed earlier, in the case of H- and O-passivated bilayer ZGNRs/AGNRs, the lower formation energies, therefore, the preferred bilayer stacking correspond to AA/AB$_\beta$ and AB$_\alpha$/AB$_\beta$, respectively. However, after Alkali intercalation, these stacking types are somehow shifted, indicating its effect on interlayer van der Waals interaction between layers, as shown in Fig.3. This is different from the case of Alkali intercalated graphite/layered graphene, in which the stable AB staking is shifted to AA stacking after intercalation. Based on the calculation result, the optimal position of alkali atoms is situated at the hollow site of one layer's honeycomb lattice. There are four main types of intercalations are investigated, namely fully-intercalation (max), half-intercalation (half), single-intercalation at the ribbon edge (min_edge), and single-intercalation at the center of ribbon (min_central), in which the max alkali distribution is similar to that of $LiC_6$ in graphite anodes[68]. In Fig.S4 the systems that are intercalated with H-GNRs are demonstrated using the same concentrations. The interlayer distance is specified in two positions indicated by the symbols $d_1$ and $d_2$ at the edge and in the middle of the ribbon's width (Fig.3). The effects of adatom passivation and intercalation on layered GNRs are described in Table 2 and Fig.4. As shown in the table, interlayer distance is distinct depending on interaction between two stacking layers and the effect of intercalant as well as passivation. The distance could be enlarged by intercalated atoms, in which sodium cases show remarkably larger distance compared to lithium systems. In each intercalant system, the distance at position $d_2$ and $d_1$ is totally different based on intercalant position. Take the central cases of ZGNRs as representative, $d_2$ is larger than $d_1$ because of intercalant locates at the middle of the ribbon (see Table 2). Moreover, the intercalant could affect the C-C bond length in planar GNRs as shown in Fig.4 leading to the strongly non-uniform environment. Furthermore, intercalant exhibits the strong bonding with oxygen (A-O bonds) at the edge and modifies C-C and C-O bonds in this region. In contrast, A-H bonds are absent in all cases, indicating that oxygen is more active than hydrogen that might come from the active 2p orbitals of oxygen atoms. Hence, alkali intercalated bilayer GNRs not only contribute to interaction between two layers but also adjust the chemical and structural environment of GNRs.

**Table 2.** Interlayer distance near the edge ($d_1$) and at the middle of the ribbon ($d_2$) of A-intercalated GNRs in both H- and O-passivated cases.

| | H-passivated | O-passivated |
|---|---|---|

|   |        | Zigzag |      |      |      | Armchair |      |      |      | Zigzag |      |      |      | Armchair |      |      |      |
|---|--------|--------|------|------|------|----------|------|------|------|--------|------|------|------|----------|------|------|------|
|   |        | (1)    | (2)  | (3)  | (4)  | (1)      | (2)  | (3)  | (4)  | (1)    | (2)  | (3)  | (4)  | (1)      | (2)  | (3)  | (4)  |
| Li | $d_1$ | 3.9 | 3.69 | 3.68 | 3.31 | 4.21 | 3.96 | 3.6 | 3.96 | 3.8 | 3.32 | 4.02 | 3.38 | 3.92 | 3.92 | 4.13 | 3.28 |
|    | $d_2$ | 3.72 | 3.65 | 3.66 | 3.76 | 4.2 | 3.84 | 3.43 | 3.89 | 4.02 | 4.04 | 3.52 | 3.93 | 3.8 | 4.0 | 3.52 | 3.92 |
| Na | $d_1$ | 4.52 | 5.38 | 4.6 | 3.57 | 4.38 | 5.04 | 4.25 | 3.66 | 4.45 | 3.96 | 3.51 | 3.46 | 5.67 | 4.37 | 3.49 | 3.71 |
|    | $d_2$ | 4.45 | 4.24 | 3.62 | 4.79 | 4.63 | 4.33 | 3.65 | 4.13 | 6.83 | 5.38 | 3.52 | 4.7 | 6.84 | 5.27 | 3.53 | 4.78 |

(1): max, (2): half, (3): min_edge, (4): min_central

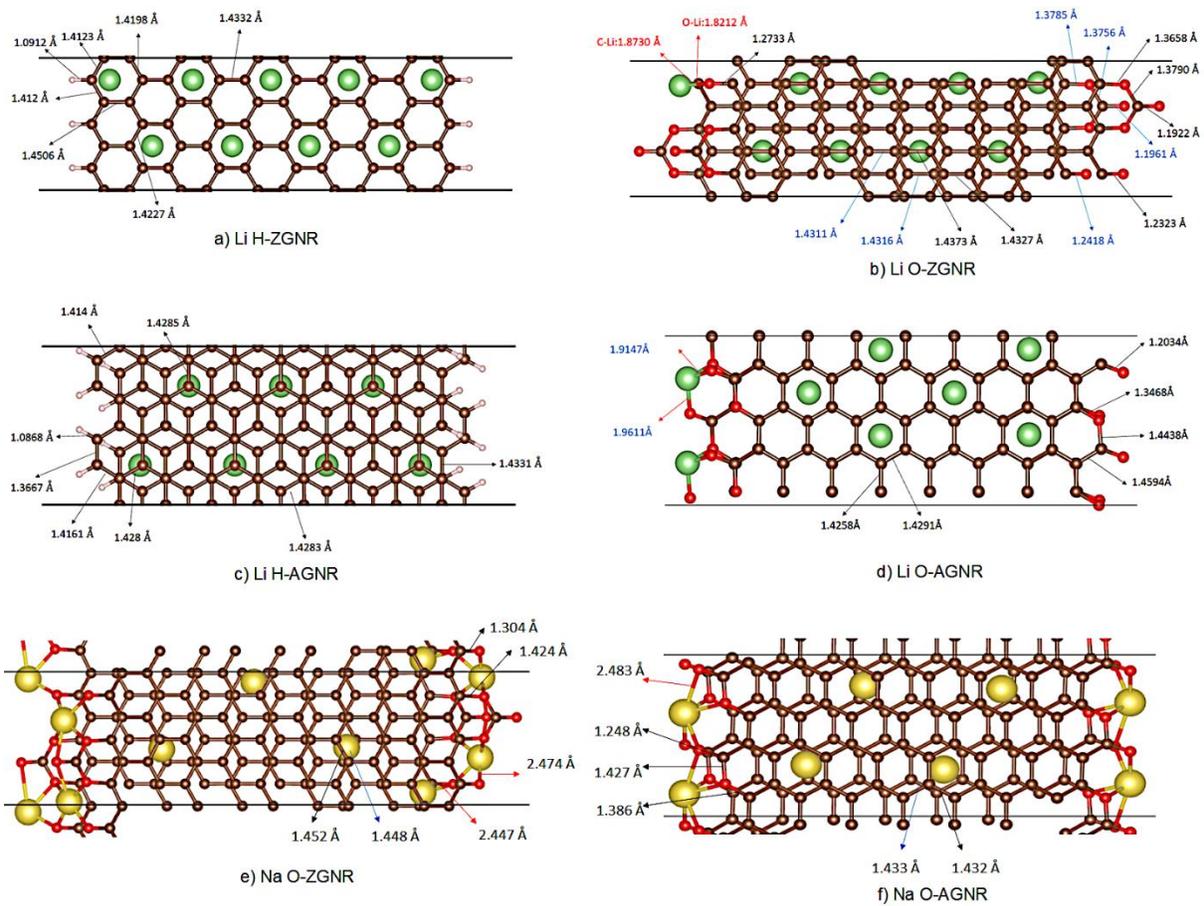

**Fig. 4** Chemical bond length (Å) of C-C, H-C, O-C and A-C bondings in the max case of Li-intercalated (a) H-ZGNR, (b) O-ZGNR, (c) H-AGNR, (d) O-AGNR and Na-intercalated e) O-ZGNR, f) O-AGNR indicates non-uniform environment after intercalating.

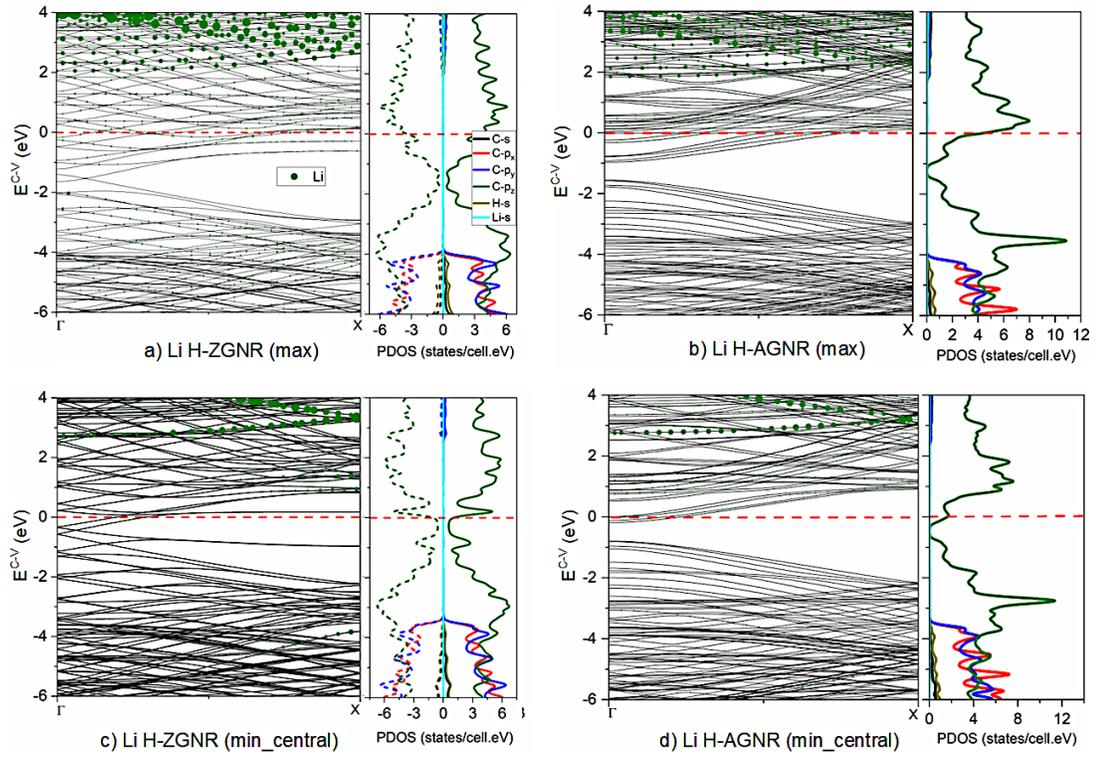

**Fig. 5** Band energy spectra of Li-intercalated bilayer (a) H-ZGNR, (b) H-AGNR in the max case and (c) H-ZGNR, (d) H-AGNR in the min_central case. Green circles represent lithium atom dominance in band structures. Solid and dash lines respectively describe spin up and down orbital density of states.

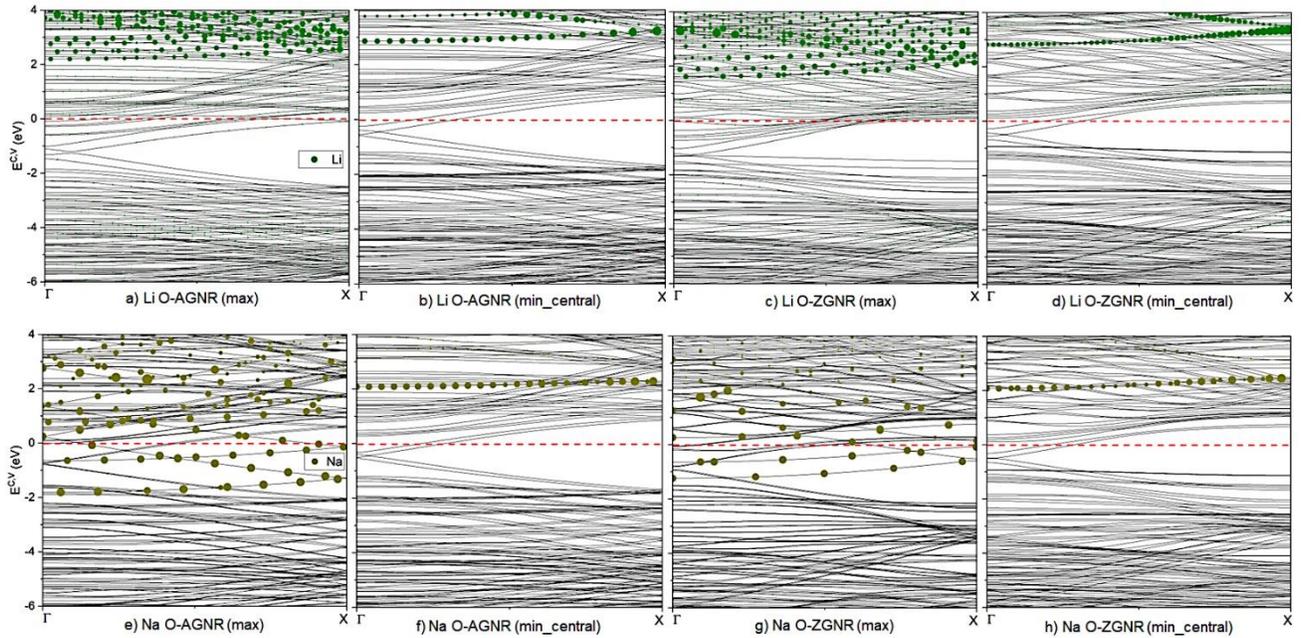

**Fig. 6** Band energy structures of (a) Li- and (b) Na-intercalated O-GNRs in the max and min_central

cases.

The alkali-intercalated GNRs exhibit the feature-rich electronic properties, being dominated by the finite-size confinement, the critical orbital hybridizations in C-C, C-A, O/H-C and O-A bonds. Band energy spectra and orbital-projected DOS provide more information about these interactions through the contribution of orbitals and atoms in the system shown in Figs.5, 6 and 7. Accordingly, the band energy structure and corresponding orbital-projected DOS of Li-intercalated H-GNRs (Fig.5), band energy spectra of Li- and Na-intercalated O-GNRs (Fig.6), and orbital-projected DOS of Li-, Na-intercalated O-GNRs (Fig.7) are considered with the max and min_central cases. Other cases are also shown in Fig.S6 in **Support information** for perspective of their characteristics. All A-intercalated GNRs exhibit metallic behavior due to the overlap between conduction and valence bands. Energy bands reveal significant blue shift with the shifting of $E_F$ toward the conduction band about 0.5 eV to 1.5 eV, which is proportional to the Li concentration. Furthermore, there exists the edge-atom-induced valence and conduction bands in which lithium atoms reveal the strong effect in conduction bands based on adatom dominance. Apart from the contribution in conduction band, sodium atoms exhibit the dominance in the valence band near the Fermi level (see Fig.6b), indicating the slightly distorted $\pi$ bonding. The main characteristics of electronic structures are directly reflected in the DOS, as shown in Fig. 7. Alkali intercalated GNRs reveal s-$2p_z$ and 2p-$2p_z$ orbital hybridizations in A-C/H-C and O-C belonging to hydrogen and oxygen passivation, respectively. Specifically, oxygen passivated cases indicate strong orbital hybridization due to 2p-$2p_z$ because of outmost 2p orbitals of oxygen (see Fig.7) while this bond is absent in hydrogen cases. Moreover, A-intercalated HGNRs remain $\pi$ bond near the Fermi level due to $p_z$ orbitals with green curves (see Fig.5). On the other hand, $\sigma$ and $sp^3$ bonds based on ($p_x$, $p_y$) and (s, $p_x$, $p_y$, $p_z$) orbitals make significant contributions to the valence DOS in oxygen passivated cases of both Li and Na intercalations (see Fig.7). Hydrogen and oxygen passivation displays distinct features from each other in bilayer GNRs in both pristine and intercalated cases.

Additionally, Na-intercalated systems show higher density of states compared to Li-intercalated cases because of more charge contribution which comes from its larger atom size. Hence, the intercalant not only affects structural characteristics such as bond length and interlayer distance, but also electronic features as mentioned in the energy bands and orbital-projected DOS.

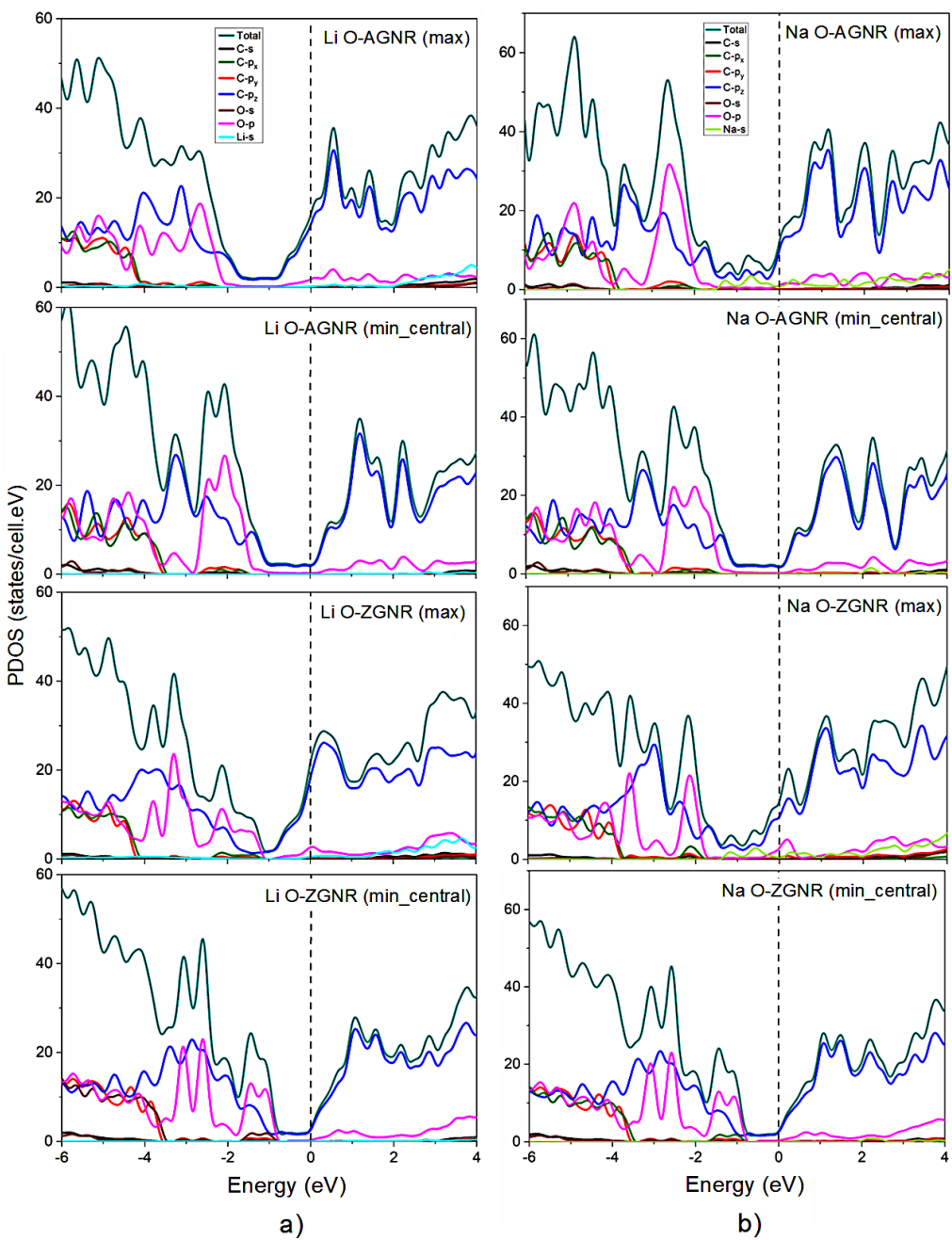

**Fig. 7** Orbital-projected DOS of (a) Li- and (b) Na-intercalated O-GNRs in the max and min_central cases.

In order to further comprehend the multi- or single-orbital hybridizations in A-C, H-C or O-C bonds, which dominate the essential properties, the spatial charge distributions have been taken into account. The charge density difference, created by subtracting the charge density of graphene and halogen atoms from that of the composite system, can provide very useful information about the chemical bondings and thus explain the dramatic changes of electronic structures. As crucial evidence for interaction, charge distribution of alkali intercalated GNRs provide more information about bonding that exhibits the coexistence and common contribution of alkali and carbon atoms (see Fig.8 and Fig.S5). In Fig.8, the charge density is exhibited in the side view for the max and min_central cases of A-intercalated O-GNRs. The charge distribution corresponding to intercalant and GNRs in cases of half and min_edge systems is also shown in Fig.S5. Obviously, the optimal intercalated GNRs display the buckled structure instead of the planar one as pristine monolayer case. This indicates that the $\sigma$- and $\pi$-bonds of carbon are somehow distorted due to the edge passivation and intercalation effect. According to the charge density, A-C and O-C bondings are revealed contribution in the region around lithium atoms indicating the van der Waals interactions and orbital hybridization in the system. Furthermore, there exist charge variations between alkali and carbon atoms, arising from the s-$2p_z$ orbital hybridization in A-C bonds and 2p-$2p_z$ orbital hybridization in O-C bonds. It is worth noting that the well-extended $\sigma$-bonding in graphene is hardly affected by alkali intercalation but dramatically affected by oxygen passivation, leading to the modifications on the carbon dominated valence bands. The $\pi$-bonds are distorted only near the adatom sites, and they behave like the normal extended states at other positions.

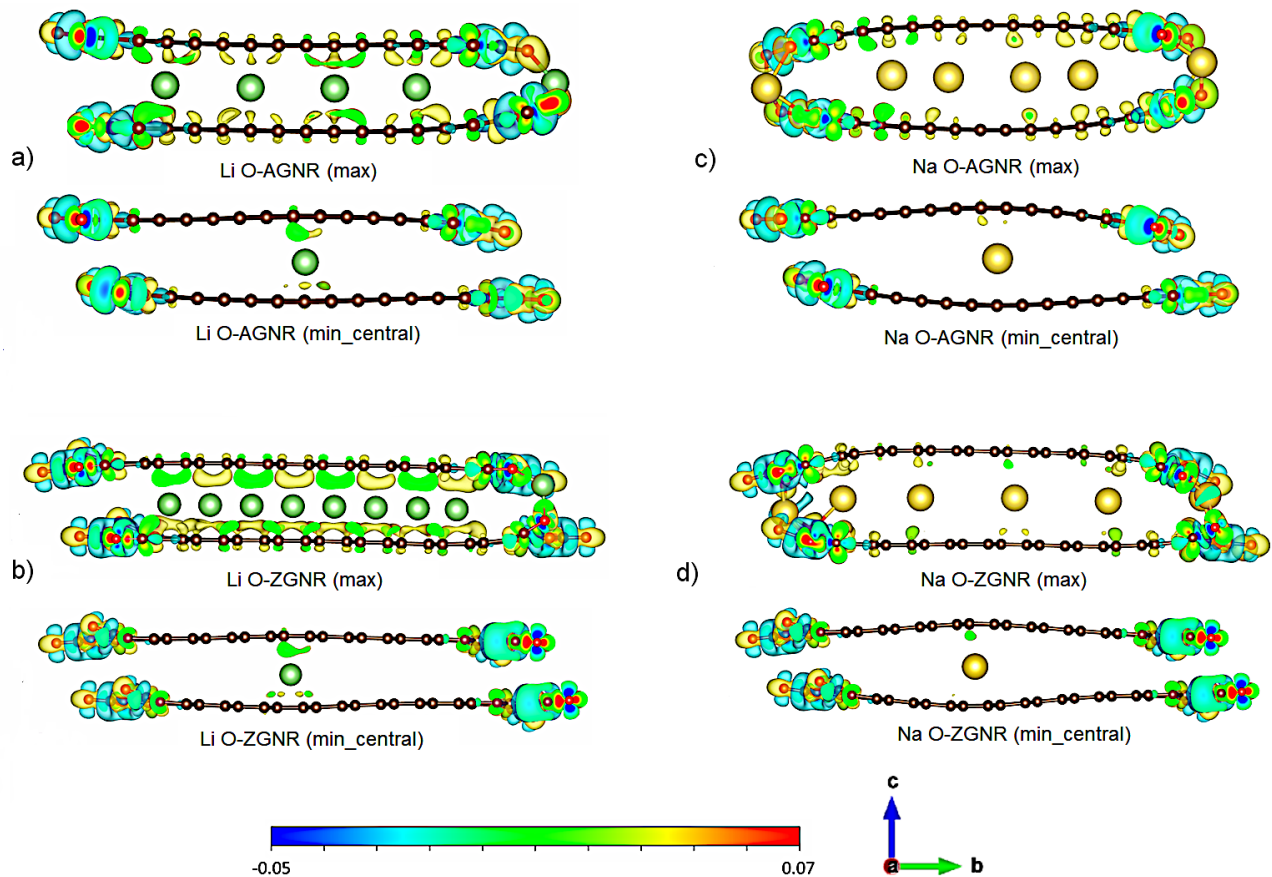

**Fig. 8** The charge density difference of Li-intercalated (a) O-AGNRs, (b) O-ZGNRs, c) Na-intercalated (c) O-AGNRs, (d) O-ZGNRs in the max and min_central cases.

Interestingly, there are significant differences in magnetic configurations between O- and H-passivated GNRs that were not reported elsewhere before. In case of non-passivation GNR, it is well-known that the ZGNR belongs to AFM, while the armchair one is non-magnetic. This AFM has zero net magnetic moment, since two edges in a zigzag system are symmetric about the nanoribbon center. Based on our spin-polarized calculations, O-passivation cases display nonmagnetic configurations for all the monolayer and bilayer structures, even with the zigzag edge. As shown previously in the electronic structures (Fig. 5, Fig 6, Fig. 7) and charge density (Fig. 8), the strong bonding between O atoms and ribbon edge's C atoms has destroyed the AFM configuration of non-passivated ZGNR. In case of H-passivation, due to the weak s-$2p_z$ orbital hybridization in H-C bonds, the magnetic configuration remains the same as non-passivation case. For bilayer GNRs, the typical spin configurations can be classified as AFM-AFM, FM-AFM, AFM-FM, and FM-FM, corresponding to their intralayer-interlayer spin arrangements. The AFM-AFM and FM-AFM models as shown in Fig.9 (a-f) are the stable magnetic configurations for all three types of stacking AA, $AB_\alpha$ and $AB_\beta$, owing to the lower ground state energies. These aforementioned differences in edge passivation and spin configurations could lead to diverse electronic properties. Under intercalation (Fig.9 (g-i)), the spin configuration is strongly distorted. Intercalation at high concentrations can destroy the surrounding spin configuration due to strong bonds between intercalated atoms and carbons as illustrated in Fig. 9(g) for the half_edge case. In contrast, the spin distribution remains at the intercalation edge but is absent at the non-intercalation edge at low intercalation (Fig. 9(h)). Similar behavior could be observed in the case of Na adatom (Fig. 9(i)). As shown by the significant difference in layer distance between the two edges of the ribbon, this could be the result of the vdW interaction effect on spin polarization. This effect of vdW on magnetic configuration is also observed in other layered systems[69].

Additionally, the main features of band structures and DOS could be examined by angle-resolved photoemission spectroscopy (ARPES) and scanning tunneling spectroscopy (STS) measurements. ARPES is an effective tool in identifying the diverse band structures of graphene-based systems, e.g., the 1D parabolic valence bands near the Γ point accompanied with band gaps and distinct energy spacings of AGNRs in the presence/absence of edge passivation[70, 71]. The greatly modified band structures of the Ti-absorbed graphene[72] or the red shift of 1.0-1.5 eV in the pi bands (n-type doping)[73] has been confirmed by using high-resolution ARPES. On the other hand, STS measurements of the dI/dV spectra, which is proportional to the local DOS, has confirmed the asymmetric peaks of 1D parabolic bands of GNRs[74, 75] and the opening of confinement energy gap[70, 76].

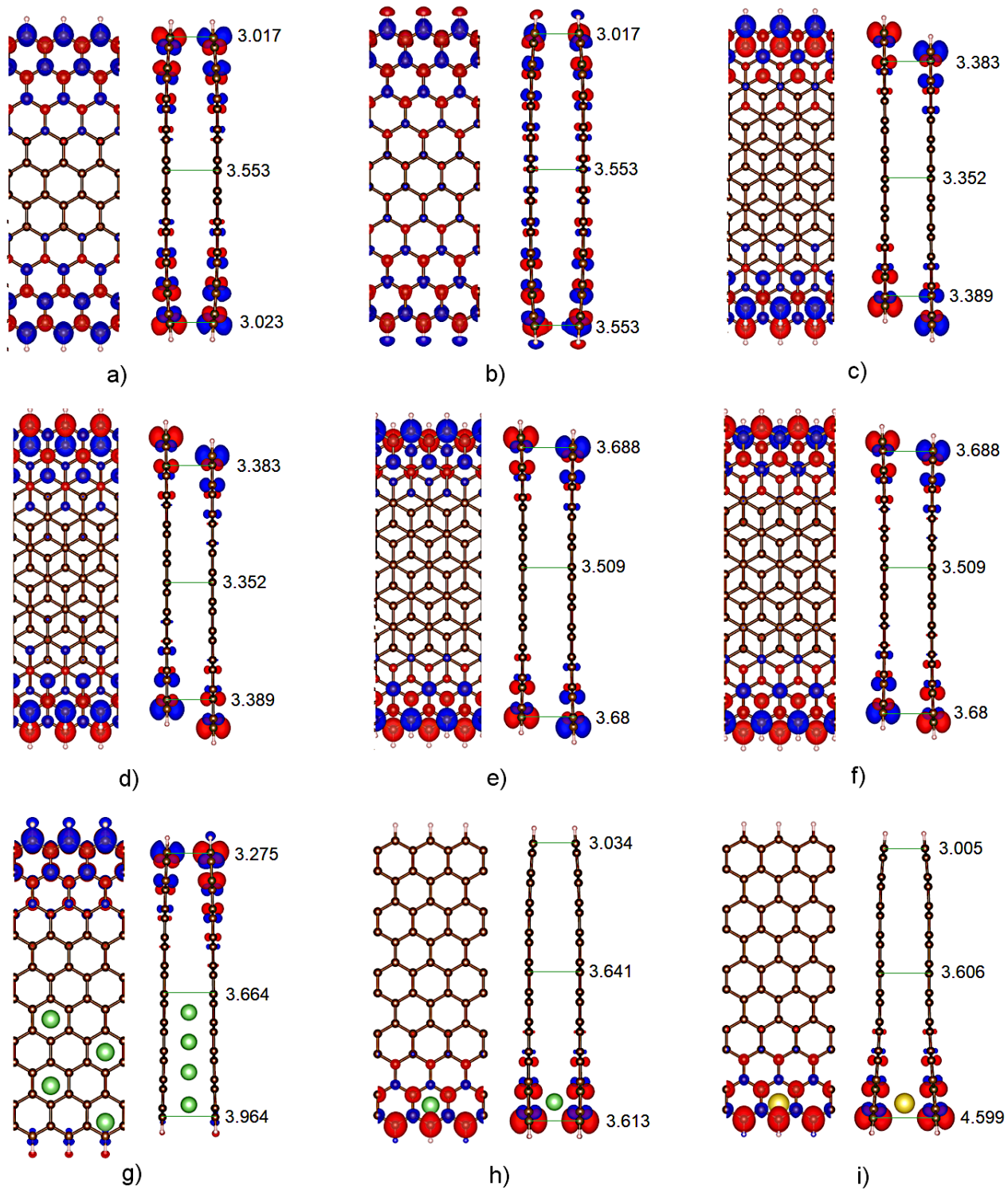

**Fig. 9** The top and side views of spin distribution considering various magnetic configurations including (a) FM-AFM, (b) AFM-AFM in AA-stacked, (c) FM-AFM, (d) AFM-AFM in AB$_\alpha$-stacked, (e) FM-AFM,( (f) AFM-AFM in AB$_\beta$-stacked H-ZGNR, Li intercalated AA H-ZGNR in g) half , (h)

min_edge cases, and (i) Na intercalated AA H-ZGNR in min_edge case. Red and blue circles describe spin up and spin down distribution. The interlayer distances at two edges and central of the ribbon were also displayed.

With regard to applications, intercalated systems are expected to play a role in the next and modern generation of materials which are technology nodes[4, 56, 68] including anodes, cathodes, electrolytes, and interconnects. Among them, interconnect performance of GNRs has become an emergent and fascinating structure in integrated circuits because of their limitations of speed consumption. Instead of using Cu in interconnects, Li intercalation exhibits an improvement in the conductivity along the c axis of GNRs[42]. Thus, Li-intercalated GNRs promise as a potential candidate for interconnect applications[41, 42], suggesting that alkali intercalated systems could support this point. Besides, one of the most appealing properties of 2D materials is superconductivity which has been applied in recent material science. However, a superconducting state in pristine monolayer graphene has not been reported, leading to a manipulation in atomic structure to induce this feature in graphene or graphene-based systems. Hence, the exhibited metallic character in our intercalated GNRs systems could propose a phonon-mediated superconducting state[50] for forming alkali-intercalated GNRs superconductor. Generally, this work is expected that our findings can stimulate further experimental studies on essential and state-of-the-art properties of intercalated bilayer GNRs for verifying the results and predictions herein.

**Concluding Remarks**

In this work, layered GNR structures with hydrogen and oxygen passivation edges have been optimized by means of first-principles calculations. A comparison between two passivations was also considered in order to verify significant and distinct features, in which H-passivated cases exhibited magnetic configuration while the non-magnetic behavior was found in all O-passivated systems. AFM-AFM and FM-AFM configurations display stable magnetic structures in bilayer and intercalated H-GNRs, in which the significant effect of intercalation and vdW interaction are observed. Moreover, the alkali-metals-intercalated layered GNRs can induce metallic behaviors, indicating higher carrier density compared to the semiconducting pristine GNR ones. This feature satisfies the main point to apply for lithium or sodium batteries based on the high carrier density. Furthermore, our systems suggesting the application in interconnect materials, or further research in superconducting states due to their metallic behavior. In each intercalated case, intercalant displays different features, resulted in interlayer distance, atom dominance in band energy and orbital distribution in DOS, in which sodium shows the stronger distribution in band energy and enlargement interlayer distance, as well as higher density of states. There are feature-rich geometric and electronic properties, being dominated by the finite-size confinement, edge passivation, Van der Waal interactions, and the critical orbital hybridizations in C-C, C-A, O-C, H-C and O-A bonds. The predicted geometric structures, electronic properties, and magnetic configurations are worthy of further experimental examinations. Hence. the present work should serve as a first step toward a further investigation into other necessary properties of alkali-metals-intercalated GNRs for fabrication and potential devices.

**Author contributions**

Thi My Duyen Huynh: conceptualization, model building, methodology, software, validation, original draft preparation, writing-reviewing and editing.

Guo-Song Hung: model building, data curation, methodology, software, visualization, original draft preparation.

Godfrey Gumbs: conceptualization, methodology, validation, reviewing and editing.

Ngoc Thanh Thuy Tran: conceptualization, model building, data curation, visualization, methodology, validation, reviewing and editing, supervision.

**Conflicts of interest**

There are no conflicts to declare.

**Acknowledgments**

This work was supported by the Taiwan Ministry of Science and Technology (MOST) under the project 108-2112-M-006-022-MY3 and the Hierarchical Green-Energy Materials (Hi-GEM) Research Center, National Cheng Kung University, Taiwan. G.G. would like to acknowledge the support from the Air Force Research Laboratory (AFRL) through Grant No. FA9453-21-1-0046